\documentstyle[aps,twocolumn,eqsecnum,graphicx]{revtex}

\begin{document}
%\bibliographystyle{prsty}

%----------------------User's Commands----------------------------
 \newcommand{\mytitle}[1]{
 \twocolumn[\hsize\textwidth\columnwidth\hsize
 \csname@twocolumnfalse\endcsname #1 \vskip2pc]}
 
 \newcommand{\mybbox}[1]{\mbox{\boldmath$#1$}}
%------------------------------------------------------------------------

\mytitle{
\title{Berry Phases of Classical Trajectories in the Presence of Hard-Wall 
Boundaries}
\author{Hsiu-Hau Lin and Tzay-Ming Hong}
\address{Department of Physics, 
National Tsing-Hua University,
Hsinchu 300, Taiwan, Republic of China}
\date{\today}
\maketitle

\begin{abstract}
We study the quantum propagator in the semiclassical limit
with hard-wall potentials.  We show that, upon each
reflection by the hard wall, a Berry phase $\pi$ is
accumulated and leads to interferences between different
classical trajectories.  Including the Berry phase caused by
the hard walls, the modified Van Vleck's formula is derived. 
We also discuss the close relations among quantum
statistics, discrete gauge symmetry, and hard-wall
constraints.  Most of all, we formulate a new quantization
rule that applies to {\it both} smooth and sharp boundary
potentials.  It provides an easy way to compute quantized
energies in the semiclassical limit and is extremely useful
for many physical systems.
\end{abstract}
}

\section{Introduction}

Path integral provides an alternative approach to formulate quantum
mechanics\cite{Feynman65,Kleinert95}.
The quantum propagator $G(x,x^{\prime };T)$, that is the key
quantity in quantum mechanics, is shown to equal the
summation over all possible paths with the same end points.
In the semiclassical limit, the dominant contribution comes
from classical trajectories and fluctuations around 
them\cite{Sepulveda92,Manning96,Stock97}.
Within the stationary phase approximation including
fluctuations up to the quadratic order, the quantum
propagator can be approximated by the Van Vleck's
formula\cite{VanVleck28}.
In general, there would be many classical trajectories that
satisfies the same boundary conditions, and the Berry phase
interferences between them are
important\cite{Berry72,Gutzwiller90}.

The interference effects among the classical trajectories
were not included until Gutzwiller's work that enables the
earlier short-time result of Van Vleck to be extended past
the first conjugate point\cite{Gutzwiller71}.
By Morse's theorem, the second variation, considered as
quadratic fluctuations around a given trajectory from
$x^{\prime }$ to $x$ in time $T$, has as many negative
eigenvalues as there are conjugate (turning) points along
the trajectory.
These conjugate points give rise to a Berry phase $\nu \pi
/2$ for the trajectory, where $\nu $ is the total number of
conjugate points along the trajectory, or sometimes referred
to as the Maslov or Morse index\cite{Maslov81}.

Not only elucidating the crossover between classical and
quantum mechanics, the semiclassical limit also provides a
convenient way to calculate the bound state energy.  Instead
of solving the Schr\"{o}dinger equation directly, the bound
state energies can also be computed by the WKB
approximation\cite{Sakurai94}.
However, the method is only applicable when the confinement
potential is reasonably ``smooth'' (compared with the
relevant energy gradient we are interested in).
In this paper, we study the semiclassical limit in the
presence of hard-wall potentials that is beyond the validity
of the WKB approximation.
We found that upon each reflection by the hard wall, a Berry
phase $\pi $ accumulates and eventually leads to an
additional phase $r\pi $, where $r$ is the number of
reflection points.
In order to account for the interference effects among
classical trajectories correctly, we rederive Van Vleck's
formula with an extra Berry phase correction due to
hard-wall boundaries. 
An elegant proof by mirror projection explains the Berry
phase $\pi $ and reveals the close connections among quantum
statistics, discrete gauge symmetry, and the hard-wall
boundary.

Following the standard stationary phase approximation and
making a Legendre transformation of the time variable in the
quantum propagator to the energy variable, we are able to
generalize the Einstein-Brillouin-Keller (EBK) quantization
rule\cite{Einstein17,Brillouin26,Keller58} with a hard-wall
correction term
\begin{equation}
\oint \sqrt{2m[E-V(x)]}dx=2\pi (n+\frac \nu 4+\frac r2),  \label{NewEBK}
\end{equation}
where $\nu $ is the number of conjugate points and $r$ is
the number of reflection points.
The usual WKB approximation is the special case $\nu =2$ and
$r=0$.
The modified EBK quantization rule in Eq.$~$(\ref{NewEBK})
relaxes the requirement of the potential smoothness in the
WKB approximation.
If the trajectory bounces from a smooth potential, it is
counted as a conjugate (turning) point.
If the trajectory gets reflected by a hard-wall like
boundary, it is counted as a reflection point.
This is of great advantage because many physical systems
including quantum dots, quantum wells, Hall bars, electronic
wave guides, etc., have both hard-wall-like potentials (from
sample edges) as well as smooth potentials (by applying
external fields) at the same time.

The paper is organized in the following way. 
In Sec.  II, we introduce the Van Vleck's formula and apply
it to two simple systems without and with hard-wall
boundaries.
We explicitly show that the Van Vleck's approximation is
incorrect in the presence of hard walls.
In Sec III, We show that the Berry phase due to the
reflection by a single hard-wall boundary is $\pi $ and
derive the modified Van Vleck's formula.
In Sec. IV, a more elegant
proof by mirror projection is presented. 
Finally, in Sec V, we derive the key result of this paper --
the modified EBK quantization rule.
We apply it to physical systems with both smooth and hard
confinement potentials and show that the modified term is
necessary to obtain the correct energy levels.
Then a brief conclusion follows.

\section{Quantum Propagator and Classical Trajectories}

In the path integral formalism\cite{Feynman65}, the quantum
propagator equals the sum over all possible paths with the
same end points,
\begin{eqnarray}
G(x,x^{\prime};T) &\equiv& \langle x| e^{-iHT}
|x^{\prime}\rangle \nonumber
\\
&=& \int_{x^{\prime}}^{x} {\cal D}[x] \exp \left( i
\int_{0}^{T} L(x,\dot x, t) dt \right),
\end{eqnarray}
where the measure ${\cal D}[x]$ denotes all possible paths
with end points $x(0)=x^{\prime}$ and $x(T)=x$.
In the semiclassical limit, the phase inside the path
integral oscillates rapidly except in the neighborhood of
the classical trajectories.
Within the stationary phase approximation including
fluctuations up to quadratic order, the propagator is
approximated by the Van Vleck's formula,
\begin{equation}
G(x,x^{\prime};T) \simeq \frac{1}{\sqrt{2\pi i}}\sum_{p} \sqrt{C_{p}} 
\exp[iA_{p} -i\nu_{p} \frac{\pi}{2} ],  \label{VanVleck}
\end{equation}
where $A_{p}(x,x^{\prime};T)$ is the action of the classical
trajectory starting from $x(0)=x^{\prime}$ and ending at
$x(T)=x$, and the subscript $p$ denotes all classical paths
with the desired end points.  The strength of the quadratic
fluctuations\cite{Gutzwiller90} around the classical
trajectory is
\begin{equation}
C_{p}= \left| - \frac{\partial^{2} A}{\partial x \partial x^{\prime}}\right|.
\end{equation}
Finally, the total number of conjugate (or turning) points
along the classical trajectory is denoted by $\nu$.  Notice
that, for each conjugate point, there is a $\pi/2$ Berry
phase associated with it.  Van Vleck's formula provides a
completely classical approximation of the quantum
propagator, in the sense that all relevant elements can be
computed from the classical trajectories.

A straightforward example of the Van Vleck's formula is a
free particle moving on the a finite ring with length $L$. 
There are infinite classical paths which satisfy the
conditions $x(0)=x^{\prime }$ and $x(T)=x$.  The total
(route) distance of each classical trajectory is
$d_n=x-x^{\prime }+nL$, where $n$ is an integer.  The action
for each trajectory is
\begin{equation}
A_n(x,x^{\prime };T)=\frac m{2T}(x-x^{\prime }+nL)^2.
\end{equation}
Taking the derivative of the action, the strength of
fluctuations around each trajectory $C_n=m/T$ is independent
of the end points and the choice of trajectories.  Since the
particle moves at constant velocity, it is obvious that
there is no conjugate point along any classical trajectory
and thus $\nu _n=0$.  Besides, because the fluctuations of
the classical trajectory of a free particle are exactly
quadratic, we expect the Van Vleck's formula to be exact for
this system,
\begin{equation}
G(x,x;T)=\sqrt{\frac m{2\pi iT}}\sum_n\exp 
\left[ i\frac{m}{2T}(x-x^{\prime }+nL)^2\right]
\end{equation}
This infinite sum can be re-written in terms of its Fourier
function with the use of Poisson summation formula in
Appendix A. Notice that
\begin{equation}
f(y)=e^{i\alpha (y+\beta )^2} \leftrightarrow 
F(p)=\sqrt{\frac{i\pi }\alpha }
e^{-ik^2/4\alpha +ik\beta }
\end{equation}
Choosing $a=L$, the summation over coordinate $y=na$ can be
turned into the summation over momentum $k_n=2n\pi /L$.  The
propagator is then
\begin{eqnarray}
G(x,x;T)=\frac 1L\sum_n\exp [ik_n(x-x^{\prime })-iE_nT],
\end{eqnarray}
where $k_n=2n\pi /L$ is the quantized momentum and
$E_n=k_n^2/2m$ is the quantized energy.  It is obvious that
the propagator $G(x,x^{\prime };T)$ calculated by the Van
Vleck's formula is exact in this case.

Let us now apply the Van Vleck's formula to another physical
system -- a free particle bouncing back and forth between
two hard walls.  We would calculate the propagator
explicitly and show that the Van Vleck's formula leads to
incorrect results.

\begin{figure}[tbp]
\centering\includegraphics[width=6cm]{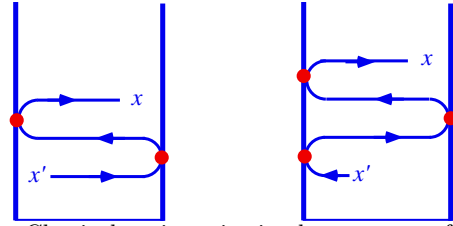}
\caption{Classical trajectories in the presence of two hard walls. On the
left is a trajectory with even reflection points $r=2,$ while the right with
odd reflection points $r=3$.}
\end{figure}

The trajectories in this problem can be classified by the
number of collisions with the hard walls, as seen in Fig. 
1.  For those trajectories that collide with the hard walls
even times, the route distance is $ d_n^e=x-x^{\prime
}+2nL$, while the distance is $d_n^o=x+x^{\prime }+2nL$ for
trajectories that collide with the walls odd times.  The
action for each trajectory can be computed straightforwardly
\begin{eqnarray}
A_n^e(x,x^{\prime };T) &=&\frac m{2T}(x-x^{\prime }+2nL)^2, \\
A_n^o(x,x^{\prime };T) &=&\frac m{2T}(x+x^{\prime }+2nL)^2.
\end{eqnarray}
Here $A^{e/o}(x,x^{\prime };T)$ denotes the action for
trajectories with even/odd reflection points.  The
fluctuations along all trajectories contribute the same
$C_n=m/T$ as in the previous example.  For an
one-dimensional motion, a conjugate point is identified as
the position where the velocity vanishes.  However, for a
free particle bouncing back and forth between two hard
walls, the velocity is constant up to a minus sign and does
not vanish at any point along the classical trajectory. 
Thus, the number of conjugate points is zero, $\nu =0$.

The propagator without any Berry
phase interference is
\begin{eqnarray}
G_{VV}(x,x^{\prime };T) &=&\sqrt{\frac m{2\pi
iT}}\sum_n\bigg\{\exp [i\frac m{2T}(x-x^{\prime }+2nL)^2]
\nonumber \\
&+&\exp [i\frac m{2T}(x+x^{\prime }+2nL)^2]\bigg\}.
\end{eqnarray}
Both infinite sums can be turned into summations over
discrete momentum again by mean of the Poisson summation
formula.  The prefactors cancel as in the previous example
and we are left with the simple result,
\begin{eqnarray}
\lefteqn{G_{VV}(x,x' ;T)=\frac 1L\sum_{n=0}^\infty \exp [-iE_nt]\times }
\nonumber \\
&\times &\bigg\{ \cos [k_n(x-x^{\prime })]+\cos [k_n(x+x^{\prime })]\bigg\},
\label{Mistake}
\end{eqnarray}
where $k_n=n\pi /L$ is the quantized momentum and
$E_n=k_n^2/2m$ is the quantized energy.  Combining two
cosines would leads to $\cos (k_nx)\cos (k_nx^{\prime }),$
while the correct form should be $\sin (k_nx)\sin
(k_nx^{\prime })$.  Indeed one can recover the exact answer
(with all prefactors right!)  if we change the sign of the
second term in Eq.~(\ref{Mistake}).  That is, only if we
assign an extra Berry phase $\pi $ to trajectories with {\it
odd} reflection points, will the modified Van Vleck's
formula become correct!

In the following section, we study the path integral
formalism in the presence of a single hard-wall boundary and
show that there is an additional Berry phase.

\section{Berry Phases Due to Hard Walls}

Consider a particle moving under the influence of a regular
potential $V(x)$ and a hard-wall potential $V_c(x)$.  The
Hamiltonian is
\begin{equation}
H=\frac{p^2}{2m}+V(x)+V_c(x),
\end{equation}
where $V_c(x)$ is the hard-wall potential at $x=0$, 
\begin{eqnarray}
V_c(x)=\left\{ 
\begin{array}{cl}
0, & x>0; \\ 
\infty , & x<0.
\end{array}
\right. 
\end{eqnarray}
The regular potential is treated in the ordinary way while
the hard-wall one is viewed as the depletion of Hilbert
space.  The complete set of the Hilbert space is now
reduced,
\begin{eqnarray}
\int_0^\infty dr|r\rangle \langle r| &=&{\bf 1}, \\
\sum_{\phi =0,\pi }\int \frac{dp}{2\pi }e^{i\phi }|p\rangle
\langle e^{i\phi }p| &=&{\bf 1}.
\label{Bases}
\end{eqnarray}
It would become clear later that the phase $\phi $ is
associated with the Berry phase in the path integral. 
Slicing the time interval $T$ into infinitesimal pieces and
inserting complete sets of the coordinate space, the
propagator is
\begin{eqnarray}
G(r,r^{\prime };T) &=&\langle r|e^{-iHT}|r^{\prime }\rangle
\nonumber \\
&=&\int_0^\infty dr_n\prod_{n=0}^{N-1}\langle
r_{n+1}|e^{-i\epsilon H}|r_n\rangle ,
\end{eqnarray}
where $r_N=r$ and $r_0=r^{\prime }$ are all positive.  Each
matrix element in the product is computed by inserting the
complete set in momentum space into Eq.(\ref{Bases}),
\begin{eqnarray}
\langle r_{n+1}| &e^{-i\epsilon H}&|r_n\rangle =\int
\frac{dp_n}{2\pi }\exp \left[ -i\epsilon H_n\right]
\nonumber \\
&\times &\sum_{x_n=\pm r_n}e^{ip_n(r_{n+1}-x_n)-i\phi _n},
\end{eqnarray}
where the phase $\phi =0$ for $x_n=r_n$, and $\phi =\pi $
when $x_n=-r_n$.  Since $x_n=\pm r_n$, the two terms can be
combined and lead to the unconstraint integral over $x_n$. 
After changing the constrained variable $ r_n$ to $x_n$, it
is convenient to write the Berry phase $\phi _n$ in the
following way
\begin{equation}
\phi _n=\pi [\Theta (x_{n+1})-\Theta (x_n)].
\end{equation}
Notice that the Berry phase is zero if the path does not
pass through $x=0$ in the infinitesimal time interval $dt_n$
and $\pi $ if the path passes through.  The integral over
momentum can be carried out easily and the propagator is
\begin{equation}
G(r,r^{\prime };T)=\sum_{x^{\prime }=\pm r^{\prime }}e^{i\phi
_B}\int_{x^{\prime }}^r{\cal D}[x]\exp [i{\cal A}(r,x^{\prime };T)].
\label{Propagator}
\end{equation}
The total Berry phase $\phi _B=\pi [\Theta (r)-\Theta
(x^{\prime })]$ is a boundary term and can be pulled out of
the path integral\cite{Kleinert95}.  The paths are divided
into two topologically distinct classes.  For all possible
paths starting from $r$ to $r^{\prime }$, the Berry phase is
zero, while for those starting from $r$ to $-r^{\prime }$,
the Berry phase is $\pi $ that causes a minus sign.  The
classical trajectories among the paths can be then
classified in the same way.  Furthermore, trajectories with
end points $r$ and $r^{\prime }$ can be identified as
trajectories (in the physical half plane) with even
reflection points and those with end points $r$ and
$-r^{\prime }$ are trajectories with odd reflection points.

Therefore, in the semiclassical limit, the Van Vleck's
formula is modified with an extra Berry phase term,
\begin{equation}
G(r,r^{\prime};T) \simeq \frac{1}{\sqrt{2\pi i}}\sum_{p} \sqrt{C_{p}} \exp[i
A_{p} - i\nu_{p} \frac{\pi}{2} - i r_{p} \pi],
\end{equation}
where $r_{p}$ is the number of reflection points.  The proof
for more than one hard wall is straightforward but tedious. 
A more general and elegant proof would be given by mirror
projection in the next section.  However, I would like to
emphasize that the proof given in this section explains
clearly the origin of the minus sign -- an extra Berry phase
due to hard-wall boundaries.

\section{Mirror Projection}

In the previous section, we treat the hard-wall boundary as
depletion of the Hilbert space.  An alternative way is to
view it as a discrete ${\cal Z}_2$ gauge symmetry of the
wave function
\begin{eqnarray}
\psi (x)=-\psi (-x).
\end{eqnarray}
The minus sign is chosen here to make the wave function
vanishes at $x=0$ so that the boundary condition $\psi
(0)=0$ is always satisfied.  Since the propagator can be
written down as the summation of eigenfunctions, $
G(x,x^{\prime };T)=\sum_n\psi _n^{}(x)\psi _n^{*}(x^{\prime
})\exp [-iE_nT],$ where $\psi _n(x)$ is the eigenfunction
with eigenenergy $E_n$.  The discrete gauge symmetry of the
wave function implies that the quantum propagator has the
symmetry
\begin{eqnarray}
G(x,x^{\prime };T)=-G(x,-x^{\prime };T).  \label{GaugeSymm}
\end{eqnarray}
Now choose both $x=r$ and $x^{\prime }=r^{\prime }$ to be
positive, the propagator can also be viewed as the wave
function $G(r,r^{\prime };T)=\psi _{r^{\prime }}(r,t)$ that
satisfies the Schr\"{o}dinger equation with a delta function
source at $(x,t)=(r^{\prime },0)$.  The propagator $
G_0(r,r^{\prime };T)$ without the hard-wall boundary
satisfies exactly the same differential equation except that
the boundary condition at $x=0$ is not met.  Notice that the
mirrored propagator $\overline{G}_0(r,r^{\prime };T) =
G_0(r,-r^{\prime };T)$ satisfies the Schr\"{o}dinger
equation without the source term since the delta function
$\delta (r+r^{\prime }) = 0$ for positive coordinates. 
Therefore, the propagator that satisfies the correct
boundary condition is constructed as
\begin{eqnarray}
G(x,x^{\prime };T)=
G_0(x,x^{\prime};T)-\overline{G}_0(x,x^{\prime };T).
\end{eqnarray}
The above result is equivalent to Eq.~(\ref{Propagator}). 
It is obvious that the discrete gauge symmetry in
Eq.~(\ref{GaugeSymm}) is satisfied.  This method is just the
familiar mirror charge trick in the classical
electromagnetism.  The generalization now becomes clear.  In
the presence of more than one hard walls, the propagator is
just the sum of all path integrals from all mirror points of
$x^{\prime }$ to the final $x$, and the Berry phase
correction is either $0$ or $\pi $ depending on how many
mirror projections are needed to reach the particular mirror
point of $x^{\prime }$.  In the example of two hard walls,
there are infinite mirror points.  Each mirror projection
corresponds to a reflection of the classical trajectory. 
Thus, for those trajectories that are reflected even times,
the Berry phase is simply zero, while for those with odd
reflection points, the Berry phase is $\pi $ and results in
a relative minus sign.

Since we can solve the hard-wall boundary by discrete gauge
symmetry, we might as well go the other way around.  It is
possible to replace the quantum statistics between particles
by the hard-wall boundaries.  Let us consider the simplest
case -- two interacting particles with either bosonic or
fermionic statistics.  The discrete gauge redundancy is
\begin{equation}
\psi(x) = e^{i\phi} \psi(-x),
\end{equation}
where $\phi =0$ for bosons and $\pi$ for fermions.  The
discrete gauge symmetry is removed by imposing a hard wall
$x_{1}=x_{2}$ in the configuration space, and a Berry phase
$\phi$ accumulates upon each reflection due to the hard
wall.

\begin{figure}
\centering\includegraphics[width=7cm]{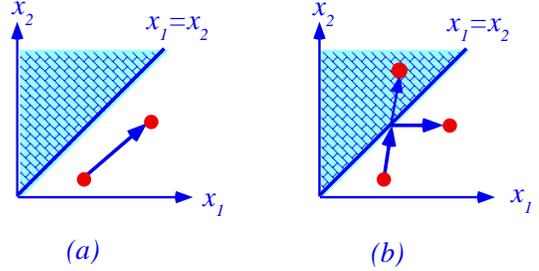}
\caption{Classical trajectories of two particles whose
quantum statistics is replaced by the equivalent hard wall
at $x_1=x_2$ in the configuration space.  In part (a), a
direct trajectory is shown and, in part (b), the shown
reflected trajectory is equivalent to exchanging two
particles which results in an extra Berry phase.}
\end{figure}

Classical trajectories are classified into two categories --
the direct path and the reflected one as shown in Figure 2. 
If we extend the reflected trajectory into the unphysical
regime inside the hard wall, as shown in Fig.  2(b), the
reflected trajectory is equivalent to an exchange between
two particles.  This approach would be useful when studying
few interacting quantum particles, e.g., two strongly
interacting bosons or fermions bouncing back and forth
between two hard walls.  In the semiclassical limit, we can
safely ignore the quantum statistics by solving all
classical trajectories inside a specific triangle in the
two-dimensional configuration space.

\section{Modified EBK Quantization Rule}

The most powerful use of Van Vleck's formula is that it
leads to the EBK quantization rule in the semiclassical
limit.  One notices that, if we set $ x=x^{\prime}$ in the
propagator and integrate over all possible $x$, it results
in the quantum partition function $Z(T) = \sum_{n}
\exp[-iE_{n}T]$.  The energy levels can then be identified
as the singularities of $Z(\omega)$ which is the Fourier
transformation of the partition function.  Within stationary
phase approximation, it can be shown that the total Berry
phase $\oint pdq - i\nu \pi/2$ (in the absence of hard-wall
boundaries) is quantized\cite{Kleinert95} and leads to the
EBK quantization rule,
\begin{equation}
\oint pdq = 2\pi (n+\frac{\nu}{4}),  \label{EBKQuanta}
\end{equation}
where $\nu$ is the number of conjugate points along the
periodic orbit.  The usual WKB approximation is the special
case with two conjugate points $\nu =2$.

\begin{figure}[tbp]
\centering\includegraphics[width=7cm]{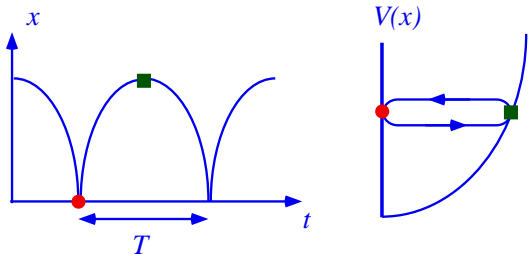}
\caption{The periodic orbit in the presence of both soft and
hard confinement potentials.  The filled circle is the
reflection point, and the filled square is the conjugate
point.  The difference between them can be seen quite
clearly by plotting the coordinate $x(t)$ as a function of
time $t $.}
\end{figure}

In the presence of the hard-wall boundaries, we need to
construct a new quantization rule that includes the extra
Berry phase.  Following a similar calculation, the new
quantization rule in Eq.(\ref{NewEBK}) is derived with the
modified term $r\pi$.  Since we have treated the hard-wall
boundary properly, the modified EBK quantization rule can be
applied to physical systems with both smooth (soft) and
sharp (hard) confinement potentials.  To illustrate this
point, let us consider a particle in a harmonic trap, but
with a hard wall boundary caused by the sample edge as shown
in Fig.  3,
\begin{equation}
V(x)=\left\{ 
\begin{array}{cl}
\frac 12kx^2, & x>0 \\ 
\infty, & x<0
\end{array}
\right. 
\end{equation}
The classical periodic orbit, shown in Fig.  3, has one
conjugate point $\nu =1$ and one reflection point $r=1$
which differs from the assumption of the WKB approximation
that there are always two conjugates points.  In the
semiclassical limit, the quantized energy is given by the
new quantization rule,
\begin{equation}
2\int_0^{x_t}\sqrt{2m(E-\frac 12kx^2)}dx=2\pi (n+\frac 14+\frac 12),
\end{equation}
where $x_t=\sqrt{2E/k}$ is the conjugate (turning) point. 
The integral is elementary and leads to the correct energy
levels
\begin{equation}
E_n=(2n+\frac 32)\omega ,  \label{OddLevels}
\end{equation}
where $n=0,1,2,\ldots $ and $\omega =\sqrt{k/m}$ is the
natural frequency.  The result can of course be verified by
the parity argument.  We emphasize that neither
Bohr-Sommerfeld quantization ($\nu =0,r=0$) nor WKB
approximation ($\nu =2,r=0$) leads to the desired result.

In many practical systems, e.g., quantum dots, quantum
wells, electronic wave guides, etc., the presence of both
soft and hard potentials is inevitable.  The modified EBK
quantization rule provides us with a convenient tool to estimate
the energy level without solving the Schr\"{o}dinger
equation directly.  For example, in the quantum well biased
by a voltage $V$ (that needs not to be small) across the
well, if the energy is below the bias voltage $V$, the
periodic orbit has one reflection point and one conjugate
point.  If the energy is above the bias voltage, we have two
reflection points.  One should not be puzzled that the
infinitely sharp hard-wall does not exist in any practical
systems and the correction due to hard-wall boundary is only
an artifact.  As long as the potential profile is sharp in
comparison to the relevant energy gradient of the interested
system, the Berry phase $\pi $ correction is reasonably
good.  On the other hand, if the potential profile is smooth
in the same sense, the Berry phase should be $\pi /2$ as in
the usual WKB approximation.

The modified EBK quantization rule can also be applied to
physical systems in higher dimensions.  Let us consider a
spherical or hemispherical qunatum dot.  We can either apply
the modified EBK formula directly to the true
three-dimensional trajectories\cite{Lin01} or apply the
formula after reducing the system to one dimension.  Here we
adapt the second approach.  After separation of variables,
the radial effective Hamiltonian of the three-dimensional
spherical (hemispherical) quantum dot becomes
one-dimensional with the effective potential
\begin{equation}
V =\left\{ \begin{array}{cl}
\frac{l(l+1)}{2mr^{2}}, &r<a\\
\infty, &r>a
\end{array}\right. ,
\end{equation}
where $l$ is the quantized angular momentum.  For the
spherical quantum dot, $l$ takes on all integer values,
while for the hemispherical dot, only odd integers are
allowed due to the flat boundary.

The classical trajectory of the electron is confined between
the hard-wall boundary at the surface and the centrifugal
potential near the origin.  Thus, there are one reflection
point $r=1$ and one conjugate point $\nu =1$.  Applying the
modified EBK quantization rule, the approximate energy
satisfies the algebraic equation,
\begin{equation}
\sqrt{(a/r_{E})^{2}-1} 
-\sec^{-1}(a/r_{E}) 
= \frac{2\pi (n+\frac34)}{\sqrt{l(l+1)}},
\end{equation}
where $r_{E} = \sqrt{l(l+1)/(2mE)}$ is the conjugate point
and $a$ is the radius of the dot.  Instead of solving the
Schr\"odinger equation directly, the energy levels can be
determined easily by the algebraic equation.  In the
semiclassical limit, the conjugate point is close to the
origin, i.e., $a/r_{E} \gg 1$.  The approximate expression
can be further simplified,
\begin{equation}
E_{n,l} \approx \frac{\pi^{2}}{2ma^{2}} 
\left(n+\frac34+\frac{l'}{2}\right)^{2},
\label{EBKEnergy}
\end{equation}
where $l' = \sqrt{l(l+1)}$.

Notice that this problem can be solved exactly by the
spherical Bessel functions.  The hard-wall boundary requires
the wave function vanishes at the surface of the sphere,
$j_{l}(\sqrt{2mE}a)=0$, that leads to quantized energy
levels.  In the same limit $a/r_{E} \gg 1$, the spherical
Bessel function is approximated by the asymptotic expansion
that leads to
\begin{equation}
E^{ex}_{n,l} \approx \frac{\pi^{2}}{2ma^{2}} \left[ n+ \frac{l}{2}\right]^{2}.
\end{equation}
The above exact result does not seem to agree with
Eq.~(\ref{EBKEnergy}) at first glance.  However, if the
angular momentum is also semiclassical ($l \gg 1$), the last
term in Eq.~(\ref{EBKEnergy}) is $l'/2 \simeq l/2 + 1/4 $ up
to $O(1/l)$ corrections.  It is then clear that both give
the same result.  We emphasize again that the agreement is
only possible when the modified term due to the hard-wall
boundary is included.

\section{Conclusions}

In this paper, we study the Berry phase of classical
trajectories due to the hard wall boundaries.  It is shown
that, upon each reflection by the hard wall, a Berry phase
$\pi $ accumulates.  We also relate the hard wall boundary
approach to the quantum statistics and the discrete gauge
symmetry.  A new quantization rule is derived from the
modified Van Vleck's formula and applied to simple examples. 
Unlike the WKB approximation that is only applicable to
smooth potential profiles, the new quantization rule
provides us with an easy way to estimate the energy levels
in the presence of both smooth and sharp confinement
potentials.

We thank Darwin Chang for fruitful discussions, especially
on the mirror projection and the discrete gauge symmetry. 
This work was supported by the National Science Council of
Taiwan, R.O.C., under Contract Nos.  89-2112-M-007-101
(HHL), 90-2112-M-007-027 (HHL), and 89-2112-M007-076 (TMH).

\appendix 

\section{Poisson Summation Formula}

Poisson summation formula provides a convenient way to
related two infinite summations together.  Let us consider a
physical system on a finite ring with length $L$ and lattice
constant $a$.  The total number of sites is $N=L/a$.  The
discrete version of the usual delta function is
\begin{equation}
\sum_{x=na}e^{ikx}=\left( \frac La\right) 
\sum_{G=2n\pi /a}\delta _{k,G},
\label{DeltaFunction}
\end{equation}
where $G$ is the reciprocal lattice vector.  Consider the
following summation,
\begin{equation}
\sum_nf(na)=\int \frac{dk}{2\pi }F(k)\sum_{x=na}e^{ikx},
\end{equation}
where $x_n=na$ and $F(k)$ is the Fourier transformation of
$f(x)$.  With the help of the identity in
Eq.(\ref{DeltaFunction}), the summation over coordinates is
turned into another summation over reciprocal momenta. 
Taking the thermodynamical limit $L\to \infty $, the
discrete delta functions are related to the continuous ones
by $L\delta _{k,G}=2\pi \delta (k-G)$.  Finally, we arrive
at the useful Poisson summation formula,
\begin{equation}
\sum_nf(na)=\frac 1a\sum_nF(\frac{2n\pi }a).
\end{equation}

\end{document}